\documentclass[floatfix,aps,twocolumn,prb,superscriptaddress]{revtex4-1}

\usepackage{amssymb}
\usepackage{amsmath}
\usepackage[dvips]{graphicx}
\usepackage{bm}
\usepackage{float}
\usepackage{color}

\newcommand{\Ef}{E_{\text{F}}}
\newcommand{\kt}{k_{\text{B}}T}
\newcommand{\be}{\begin{equation}}
\newcommand{\ee}{\end{equation}}
\newcommand{\bea}{\begin{eqnarray}}
\newcommand{\eea}{\end{eqnarray}}

\newcommand{\eps}{\epsilon}
\newcommand{\om}{\omega}

\begin{document}

\title{Transport through side-coupled multilevel double quantum dots in the Kondo regime}

\author{J. A. Andrade}
\affiliation{Centro At{\'o}mico Bariloche, CNEA, 8400 Bariloche, Argentina}
\affiliation{Consejo Nacional de Investigaciones Cient\'{\i}ficas y T\'ecnicas (CONICET), Argentina}
\author{Pablo S. Cornaglia}
\affiliation{Centro At{\'o}mico Bariloche and Instituto Balseiro, CNEA, 8400 Bariloche, Argentina}
\affiliation{Consejo Nacional de Investigaciones Cient\'{\i}ficas y T\'ecnicas (CONICET), Argentina}
\author{A. A. Aligia}
\affiliation{Centro At{\'o}mico Bariloche and Instituto Balseiro, CNEA, 8400 Bariloche, Argentina}
\affiliation{Consejo Nacional de Investigaciones Cient\'{\i}ficas y T\'ecnicas (CONICET), Argentina}

\begin{abstract}
We analyze the transport properties of a double quantum dot device in the side-coupled configuration. A small quantum dot (QD), having a single relevant electronic level, is coupled to source and drain electrodes. A larger QD, whose multilevel nature is considered, is tunnel-coupled to the small QD. A Fermi liquid analysis shows that the low temperature conductance of the device is determined by the total electronic occupation of the double QD. 
When the small dot is in the Kondo regime, an even number of electrons in the large dot leads to 
a conductance that reaches the unitary limit, while for an odd number of electrons a two stage Kondo effect is observed and the conductance is strongly suppressed. 
The Kondo temperature of the second stage Kondo effect is strongly affected by the multilevel 
structure  of the large QD.
For increasing level spacing, a crossover from a large Kondo temperature regime to a small Kondo temperature regime is obtained when the level spacing becomes of the order of the large Kondo temperature. 
\end{abstract}


\maketitle

\section{Introduction}
\begin{figure}[tbp]
\includegraphics[width=7.5 cm]{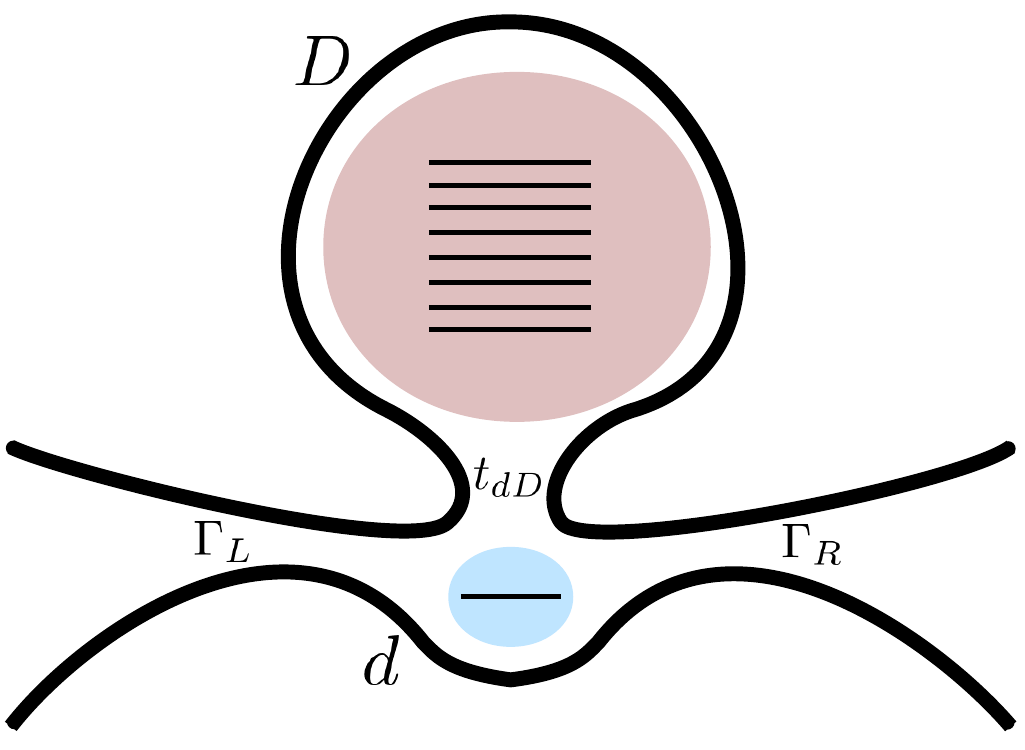}
\caption{(color online) Schematic representation of the double quantum dot device in the side-coupled configuration. QD $d$ is coupled to left ($L$) and right ($R$) metallic electrodes. The orbitals of QD $D$ are coupled to QD $d$ through a tunneling barrier.}
\label{fig:device}
\end{figure}

Double quantum dots (DQDs) laterally defined in semiconductor heterostructures are highly tunable electronic devices whose energy level spacings and charging energies can be determined by setting the size and geometry of each dot, and their coupling be tuned using metallic gates.\cite{GrabertD92,Kastner92,Beenakker1991,MesoTran97,*KouwenetalRev97,AleinerBG02,Alhassid00} 
The great tunability of the DQDs parameters, as the interdot coupling, open the possibility of applications in both classical and quantum computing,\cite{Loss1998,DasSarma2000,Koppens2006,Hayashi2003,Hanson2007} and allow a detailed analysis of the interplay between interference and correlation phenomena. \cite{Potok2007,PhysRevB.71.075305,PhysRevB.81.115316,Ferreira2011} 
The properties of these devices can be probed through transport measurements using metallic electrodes. When one QD having an odd number of electrons and a single relevant electronic level is tunnel coupled to metallic source-drain electrodes, the electric conductance increases below a characteristic temperature $T_K$ which signals the buildup of Kondo correlations. The Kondo effect is associated to the screening of the QDs magnetic moment by the Fermi sea of the metallic electrodes. In the so-called side-coupled configuration (See Fig. \ref{fig:device}), were only one of the QDs is coupled to the electrodes, the interdot coupling can compete with the Kondo effect leading to a rich variety of correlated regimes. Namely, two-stage Kondo physics 
for a small side-coupled QD having a single relevant electronic level,\cite{PhysRevB.71.075305,Ferreira2011} and a two-channel Kondo effect for a large side-coupled QD.\cite{Potok2007} 
A device with an intermediate size of the side-coupled dot has been predicted to be a realization of the Kondo box problem. \cite{PhysRevB.73.205325,PhysRevLett.82.2143,Hu2001,Simon2002,Cornaglia2003b,*Cornaglia2002a,mirage,Kaul2005,Yoo2005,Bomze2010,Kaul2006}
The properties of this type of devices have received considerable theoretical and experimental attention.
However, most theoretical studies have focused on simplified models considering a single interacting level on each dot,\cite{PhysRevB.71.075305,PhysRevB.81.115316,Zitko2006,Yoshida20093312,Cornaglia2012,Tifrea2013} a continuum of levels of the side-coupled dot,\cite{Twochannel2003,Twochannel2004} or no interactions on the large dot.\cite{Aldea2011}

Recent experimental results for a DQD device in the side-coupled configuration, \cite{Baines2012} show a change in the level structure of the DQD as a function of the interdot tunneling coupling. 
In a spectroscopic regime, where the hybridization between the small QD and the leads is smaller than the thermal energy, 
the Coulomb blockade diamonds change their structure as a function of the interdot tunneling coupling signaling a change in the underlying electronic structure associated to the multilevel nature of the QDs. 

In this article we analyze the transport properties of a DQD device in the side-coupled configuration. 
We consider the multilevel nature of the side-coupled dot and the intra- and inter-dot Coulomb interactions are properly taken 
into account. 
We obtain an exact relation between the zero-temperature conductance and the charge in the DQD device 
(assuming energy independent lead-dot couplings) which generalizes Friedel's sum rule for the double-dot device.   
We analyze the different 
strongly-correlated-electron regimes 
that occur at finite temperatures using Wilson's numerical renormalization group (NRG) and characterize the electronic transport through the device. 

In the weak coupling regime (for temperatures higher than the coupling to the leads) we use a perturbative approach, starting from the exact eigenstates for the system of two dots isolated from the leads. In the strong coupling regime, we also use a slave-boson
mean-field approximation to help the interpretation of the NRG results.  
We show that the usual even-odd asymmetry in the Coulomb blockade valleys, due to the Kondo effect, can be completely altered 
in the large interdot hopping regime. 

The effect of increasing the interdot hopping on the conductance is very different for different occupations. 
In the strong-coupling regime, we show that there is a subtle competition between the level spacing in the large quantum dot
and the effective Kondo coupling between both dots, leading to a crossover at a very small energy scale related with a 
second-stage Kondo effect.

The rest of this article is organized as follows. In Sec. \ref{sec:model} we describe the model and basic formulas for the transport calculations. In Sec. \ref{sec:wc} we present the conductance and the DQD occupation in the regime of weak-coupling between the electrodes and the DQD for the experimental parameters of Ref. [\onlinecite{Baines2012}]. In Sec. \ref{sec:exact} we present exact results for the zero temperature conductance. In Sec. \ref{sec:NRG} we present numerical results for the conductance in the Kondo regime.
In Sec.\ref{sec:qdeg} we calculate the conductance and the magnetic susceptibility for a model with two and three 
quasidegenerate levels in the side-coupled QD. We also analyze an effective model in the slave-boson mean-field approximation. 

\section{Model}\label{sec:model}
The double quantum dot device is described by the following Hamiltonian
\begin{equation} \label{eq:hamilt}
  H=H_C+ H_{t}+H_e+H_{V}+H_{\text el}\;.
\end{equation}
Here $H_C$ describes the electrostatic interaction in the constant interaction approximation\cite{Wiel2002}
\begin{eqnarray} \label{eq:coul}
H_C &=& \sum_{\ell=d,D} \frac{U_\ell}{2} (\hat{N}_{\ell}-\mathcal{N}_{\ell} )^2 \nonumber\\&+&U_{dD}(\hat{N}_D-\mathcal{N}_D )(\hat{N}_d-\mathcal{N}_d ),
\end{eqnarray}
where $\hat{N}_{d}=\sum_{\sigma} d_{d\sigma}^\dagger d_{d\sigma}$ is the occupation of the small dot (QD $d$), 
$\hat{N}_D= \sum_{\sigma,\alpha} d_{D\alpha\sigma}^\dagger d_{D\alpha\sigma}$ is the occupation of the large dot (QD $D$),
$\mathcal{N}_{\ell}=C_{g\ell} V_{g\ell}/U_\ell$, $C_{g\ell}$ is the capacitance of dot $\ell$ with its corresponding gate electrode, $U_\ell$ is the charging energy and $U_{dD}$ is given by the QDs mutual capacitance.\cite{Kastner92,KouwenetalRev97,AleinerBG02,Alhassid00} 
\begin{eqnarray}
H_{t}=\sum_{\sigma,\alpha}t_{dD}^{\alpha}
\left(d^\dagger_{d\sigma} d_{D\alpha\sigma} + h.c\right),
\end{eqnarray}
describes the tunneling coupling between the different orbitals on the side-coupled QD $D$ and the smaller QD $d$ having a single relevant electronic level. To describe the energy level splitting on QD $D$ we include a single electron energy term:
\be
H_e =
\sum_{\sigma,\alpha}\tilde{\epsilon}_{D\alpha} d_{D\alpha\sigma}^\dagger d_{D\alpha\sigma}.
\label{eq:single}
\ee 
Finally,
\be
H_V =\sum_{\nu=L,R} \sum_{k,\sigma}V_{k\nu}\left[c_{\nu k\sigma}^\dagger d_{d\sigma} + h.c. \right],
\ee\\
describes the coupling between QD $d$ and the left ($L$) and right ($R$) electrodes, 
which are modeled by two non-interacting Fermi gases:
\be
H_{\text el} = \sum_{\nu, k,\sigma} \epsilon_k c_{\nu k\sigma}^\dagger c_{\nu k\sigma}.
\ee

The conductance through the system is given by\cite{Meir1992,Pastawski1992,Beenakker1991}
\be\label{eq:Gsimp}
G=\frac{e^2}{\hbar}\frac{\Gamma_R \Gamma_L}{\Gamma_R+\Gamma_L}\sum_\sigma\int d\epsilon\left[-\frac{\partial f(\epsilon)}{\partial \epsilon}\right]A_{d\sigma}(\epsilon).
\ee
Here $A_d(\epsilon)$ is the spectral density of the small QD and we have assumed proportional ($\Gamma_L\propto \Gamma_R$) and  energy independent dot-lead hybridization functions:
\be \label{eq:gamma}
{ \Gamma}_{L(R)}= 2 \pi \rho_{L(R)}(\Ef) V^*_{L(R)}(\Ef)V_{L(R)}(\Ef).
\ee
where $\Ef=0$ is the Fermi energy of the electrodes, $\rho_{L(R)}(\epsilon)$ is the electronic density
of states of the left (right) electrode, and $V_{L(R)}(\eps)$ equals $V_{kL(R)}$ for $\eps=\eps_k$. 

In the zero temperature limit: $-\frac{\partial f(\epsilon)}{\partial \epsilon}\to\delta(\epsilon)$, and the conductance is proportional to the spectral density at the Fermi level $\sum_\sigma A_{d\sigma}(0)$. As we show in Sec. \ref{sec:exact} assuming a Fermi liquid ground state, $A_{d\sigma}(0)$ can be written as a function of the total electronic occupation of the DQD.

In the regime of weak lead-QD couplings or high temperatures $\Gamma=\Gamma_L + \Gamma_R \ll\kt$, 
we can calculate the conductance through the system, to lowest order in $\Gamma/k_BT$, replacing in Eq. (\ref{eq:Gsimp}) the exact spectral density $A_{d\sigma}(\epsilon)$ of the isolated DQD:
\bea
A_{d\sigma}(\epsilon)&=&\frac{1}{Z}\sum_{i,j} (e^{-\beta E_i}+e^{-\beta E_j}) |\langle\Psi_j|d_{d\sigma}^\dagger|\Psi_i\rangle|^2\nonumber\\&\times&
 \delta[\epsilon-(E_j-E_i)],
\eea\\

where $|\Psi_i\rangle$ and $E_i$ are the exact eigenfunctions and eigenenergies of the DQD, and $Z=\sum_i e^{-\beta E_i}$ is the partition function.
Replacing this in Eq. (7) 
\bea \label{eq:condf}
G&=&\frac{e^2}{\hbar}\frac{\Gamma}{\kt} \sum_{i,j} (P_i+P_j) f(E_i-E_j)f(E_j-E_i)\nonumber\\&\times&\sum_n |\langle\Psi_j|d_{d\sigma}^\dagger|\Psi_i\rangle|^2 
\eea
where $P_i=e^{-\beta E_i}/Z$.

The spectral function of the DQD can be calculated non-perturbatively as a function of the temperature using Wilson's Numerical Renormalization Group (NRG).\cite{Krishnamurthy1980,Wilson1975,BullaRMP2008} This allows a calculation of the conductance in the full range of temperatures with a high precision. 

\section{Perturbative results in the weak-coupling regime}\label{sec:wc}
In this section we analyze the conductance through the DQD in a regime of weak coupling to the electrodes in which the thermal energy $k_BT$ is larger than the hybridization energy $\Gamma$. In this regime, $\Gamma$ can be treated perturbatively and the electrodes serve as a spectroscopic probe of the DQD in transport measurements. 
Conductance maps are generated sweeping the gate voltages of each QD (parametrized here by $\mathcal{N}_d$ and $\mathcal{N}_D$) which modify the total charge on the DQD and the distribution of charge between the QDs. 
For a DQD in the spectroscopic regime, a hexagonal structure is expected in the conductance maps with conductance peaks occurring at gate voltages such that there is a charge degeneracy on the DQD which allows charge fluctuations between the DQD and the leads. \cite{Wiel2002} 
For the side-coupled configuration, the conductance is only sensitive to charge fluctuations on QD $d$ which is connected to the electrodes. 
As it can be inferred from Eq. (\ref{eq:condf}), to obtain a large conductance, at least two levels of the DQD differing in their charge by one electron need to be quasidegenerate ($|E_i-E_j|\lesssim \kt$ so that the product of Fermi functions is not exponentially suppressed). Furthermore, the electron must be fluctuating thermally in and out of QD $d$ (in order for the matrix element $|\langle\Psi_j|d_{d\sigma}^\dagger|\Psi_i\rangle|^2$ to be sizable).   

\begin{figure}[tbp]
\includegraphics[width=8.5 cm]{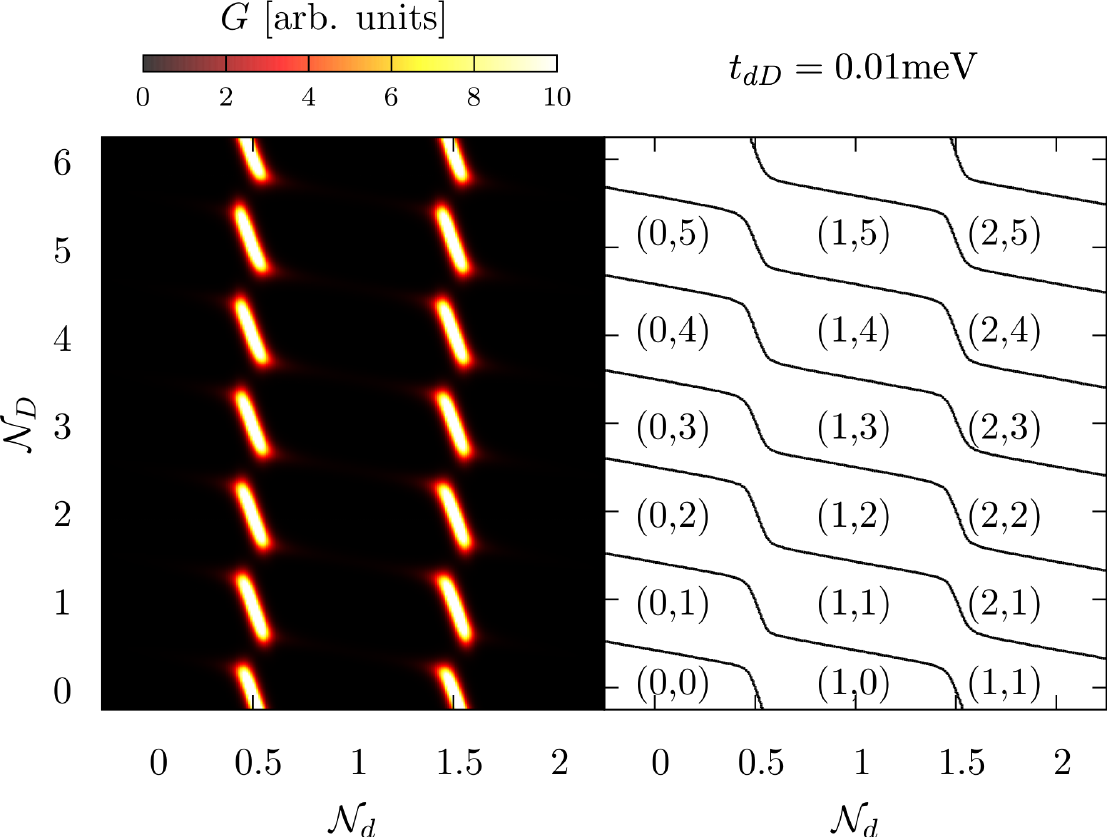}
\caption{(color online) Left panel: Conductance map for a double quantum dot device in the side-coupled configuration. The interdot coupling is $t_{dD}=0.01$meV.  Other parameters are $U_{D}=0.25$meV, $U_{d}=0.7$meV, $U_{dD}=0.1$meV, $\delta=0.02$meV, and $k_B T=0.01$meV. Right panel: Total charge of an isolated DQD for the same parameters and gate voltages as the conductance map on the left panel. The lines indicate a change by one electron in the charge of the ground state of isolated DQD. The DQD is empty at the lower left corner of the figure and is occupied with eight electrons at the upper right corner. 
The approximate occupation of each QD is indicated between parentheses as ($\langle\hat{N}_d\rangle$,$\langle\hat{N}_D\rangle$).}
\label{cm1}
\end{figure}

In what follows we focus our analysis on a DQD model having three levels on the side-coupled QD with level energies $\tilde{\epsilon}_{D1}=-\delta$, $\tilde{\epsilon}_{D2}=0$, and $\tilde{\epsilon}_{D3}=\delta$. This simplified model presents, in the weak-coupling regime, many of the features observed in systems having a larger number of levels on the side-coupled QD (see Ref. [\onlinecite{Baines2012}]). To further simplify the discussion of the results, we consider $t_{dD}^\alpha$ to be level independent and drop the level index $\alpha$. 

\begin{figure}[tbp]
\includegraphics[width=8.5 cm]{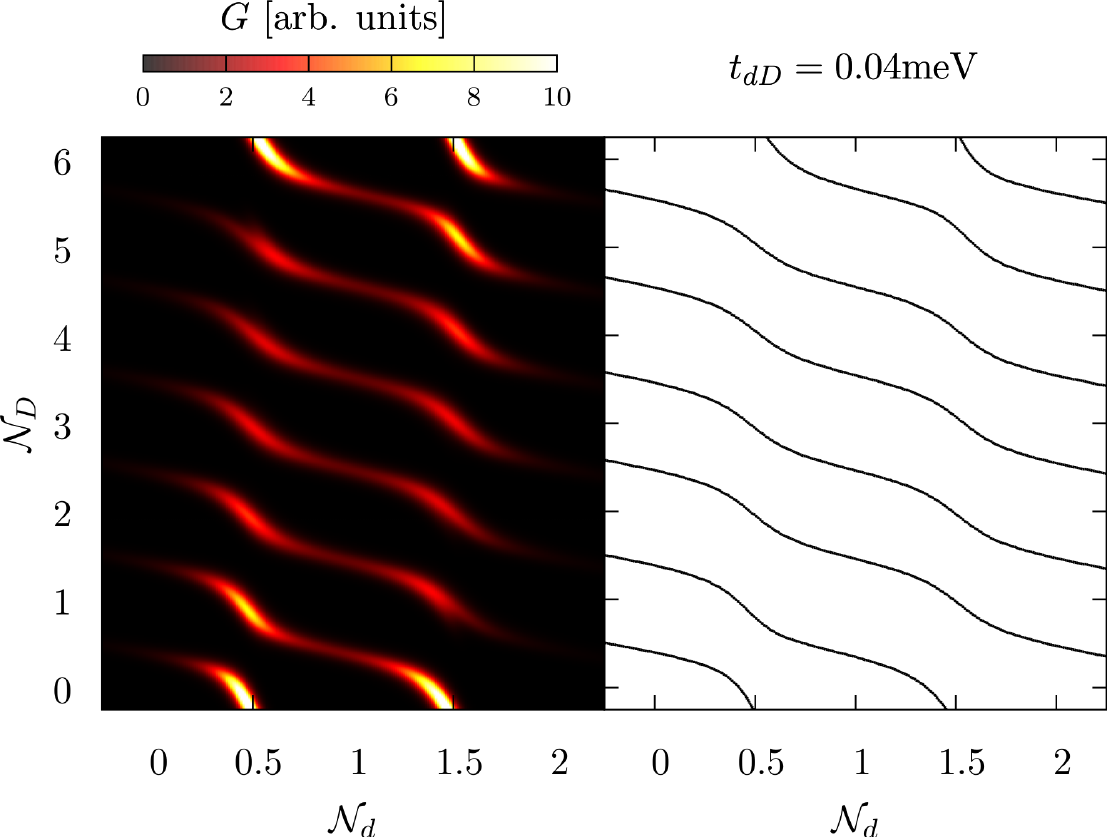}
\caption{ (color online) Left panel: Conductance map for $t_{dD}=0.04$meV. Right panel: the lines indicate a charge degeneracy on the isolated DQD, the charge in the DQD changes by one at each line and increases from zero at the lower left corner to eight at the upper right corner.  Other parameters are as in Fig. \ref{cm1}.}
\label{cm2}
\end{figure}

\begin{figure}[tbp]
\includegraphics[width=8.5 cm]{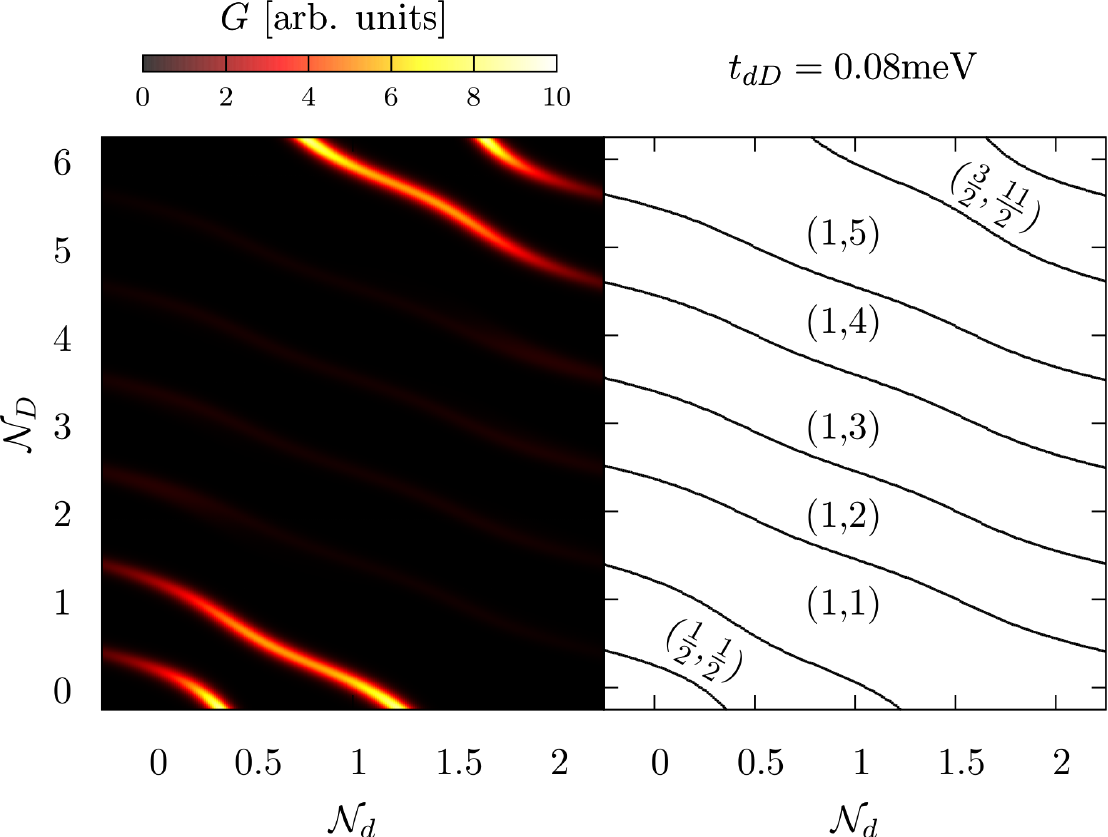}
\caption{(color online) Left panel: Conductance map for $t_{dD}=0.08$meV. Right panel: the lines indicate the charge degeneracy points of the DQD, the charge in the DQD changes by one at each line and increases from zero at the lower left corner to eight at the upper right corner. The approximate occupation of each QD is indicated between parentheses on different regions of the map. Other parameters are as in Fig. \ref{cm1}.}
\label{cm3}
\end{figure}

We calculate conductance maps of the DQD in the spectroscopic regime using Ec. (\ref{eq:condf}). The Hamiltonian of the isolated DQD is diagonalized to obtain its eigenvectors and eigenvalues for each value of $\mathcal{N}_d$ and $\mathcal{N}_D$. In Figs. \ref{cm1}, \ref{cm2}, and \ref{cm3} we present the results for a set of parameters obtained from the experiment of Baines {\it et al.},\cite{Baines2012} and different values of the interdot tunnel coupling $t_{dD}$. In the regime of weak interdot coupling ($t_{dD}<\delta$), there is little mixing of the states between QDs (except when there is a degeneracy of the energy levels of the two QDs) and the charge on each QD is generally well defined. Due to the topology of the device, a peak in the conductance is expected at the 
lines of charge degeneracy in QD $d$, which is the one coupled to the electrodes. 
This is observed on the left panel of Fig. \ref{cm1} where segments of high conductance are obtained at the charge degeneracy lines of the DQD which coincide with those of QD $d$. As it can be seen in the right panel of Fig. \ref{cm1}, these segments of high conductance are associated to a change in the charge of QD $d$.
In this figure, the approximate occupation of each QD at the different regions of the map is indicated between parenthesis ($\langle\hat{N}_d\rangle$,$\langle\hat{N}_D\rangle$).
The charging of the large dot is accompanied by much weaker peaks in the conductance due to a small mixing between the states of the two QDs.   

For an intermediate interdot coupling ($t_{dD}\sim \delta$,
see Fig. \ref{cm2}), the DQD states are in a molecular regime where the charge on each dot is not well defined as all states of the DQD involve a large orbital mixing between the QDs. This is reflected on the conductance map that presents maxima at the charge degeneracy points of the DQD where the total charge in the DQD changes (as it is indicated in the right panel of the figure). The QDs lose their identity and behave as a single QD with an effective charging energy and coupling to the leads.

For larger values of the interdot coupling $t_{dD}>\delta$ the system gradually enters a regime in which some of the wave functions present a strong mixing between the QDs and other are mostly localized on one of the QD.\cite{Baines2012} For our simplified model with a single level on QD $d$ and three levels on the side-coupled QD, one energy level is associated to a bonding state between the level on QD $d$ and a symmetric combination of levels on QD $D$, and another level is associated to the corresponding antibonding state. 
These states have approximately $1/2$ of the weight on each QD. The two remaining states of the DQD have most of their weight on the side-coupled QD. This particular electronic structure of the DQD is reflected on the conductance of the device as it can be observed in Fig. \ref{cm3}. The highest conductance peaks are obtained when the bonding (ground state) and anti-bonding (highest energy) states change their occupation and much lower peaks are obtained when the two states mainly localized on QD $D$ get charged.

This peculiar structure of the wave functions in the regime of strong interdot tunnel-coupling can be easily understood in the non-interacting limit or the high capacitance limit where $U_D=U_d=U_{dD}$, see Ref. [\onlinecite{Baines2012}]. It appears in the general case considering more levels on each QD. 
In the latter case, some states of the DQD are strongly hybridized between the two QDs, being a combination of several orbitals from each QD. The remaining states of the DQD are mostly localized on QD $d$ or in QD $D$. The value of the hopping $t_{dD}$ at which the crossover takes place depends on the level spacing $\delta$ and on the relative value of the intradot ($U_d$, $U_D$) and interdot ($U_{dD}$) Coulomb interactions. For $U_D=U_d=U_{dD}$ the interaction energy given by Eq. (\ref{eq:coul}) depends only on the total number of electrons on the DQD and is independent of the structure of the electronic wave-functions. For $U_D,U_d> U_{dD}$, states strongly hybridized between the two QDs have a larger interaction energy than those having a well defined number of electrons on each QD.

In the next sections, we analyze the low temperature conductance in the different tunnel coupling regimes.
As we will show in the next section, the conductance maps in the zero temperature limit are only determined by the total occupation $N$ of the DQD and a parameter that measures the asymmetry of the coupling of the small QD to the left and right electrodes. This allows a direct estimation of the zero-temperature conductance from the charging diagrams presented on the right panels of Figs. \ref{cm1}, \ref{cm2}, and \ref{cm3}. 

\section{Exact results at zero temperature}\label{sec:exact}
Assuming that the ground state of the system is a Fermi liquid, we show in Appendix \ref{sec:SFR} that the zero temperature spectral density of QD $d$ is given at the Fermi energy by 
\be \label{eq:AzeroT}
A_{d\sigma}(0)=\frac{2}{\pi \Gamma}\sin^2(\pi N_\sigma),
\ee
where 
\be
N_\sigma=-\frac{1}{\pi} \text{Im} \text{Tr}\int_{-\infty}^0 {\bf G}_\sigma(\om) d\om,
\ee
is the total charge per spin on the DQD and ${\bf G}_\sigma(\omega)$ is the local zero temperature Green's function of the DQD (see Appendix \ref{sec:SFR}). 
Replacing Eq. (\ref{eq:AzeroT}) in Eq. (\ref{eq:Gsimp}) we obtain
\be \label{eq:GFL}
G(T=0)=\widetilde{\alpha}\sum_\sigma\frac{e^2}{h}\sin^2(\pi N_\sigma),
\ee
where $\widetilde{\alpha}=\frac{4\Gamma_L \Gamma_R}{(\Gamma_L+\Gamma_R)^2}$. 

For a symmetric coupling of the small QD to the left and right electrodes ($\Gamma_L=\Gamma_R$) we have $\widetilde{\alpha}=1$ and the zero temperature conductance reaches the quantum of conductance when the number of electrons $N$ in the DQD is odd ($N_\uparrow=N_\downarrow=N/2$). Conversely, for an even number of electrons in the DQD the conductance vanishes. 
This important result indicates that the maximum amplitude of the conductance is given only by the asymmetry of the coupling to the left and right leads, and the remaining dependence on the parameters, including charging energies, level energies and tunnel couplings enters only through the value of the occupation in the DQD. For fixed $\widetilde{\alpha}$, 
all sets of parameters that lead to a given occupation in the DQD will lead to 
the same value for the conductance at zero temperature.  

The conductance maps analyzed in the previous section are expected to be completely modified in the zero temperature limit. Irrespectively of the charge distribution inside the DQD, the conductance will be $\sim \widetilde{\alpha} 2e^2/h$ in the valleys where the total number of electrons is an odd integer and very small in those where the total number of electrons is even. In the Kondo regime, where the hybridization $\Gamma$ is smaller than the local interactions $U_d$ and $U_D$, the occupation as a function of the gate voltages is well approximated by the occupation of the isolated DQD away from the charge degeneracy points. At the charge degeneracy points, the total charge in the DQD is $N=k +1/2$, where $k$ is an integer and the conductance at $T=0$ is $e^2/h$. Based on the calculation of the total occupation of the DQD as presented in the right panels of Figs. \ref{cm1}, \ref{cm2}, and \ref{cm3}, we expect the conductance maps at $T=0$ to consist of diagonal stripes of large conductance (delimited by the charge degeneracy lines) alternating with stripes of low conductance. 

In the next section we show that the temperatures at which the Fermi liquid behavior sets in and the above equation is satisfied can, in some cases, be extremely low. At intermediate temperatures, the conductance can be very different to the value expected in the high temperature and the low temperature regimes. Depending on the value of the gate voltages (i.e. at different regions of the conductance maps), the system can be in the spectroscopic regime or the low temperature regime for a given set of parameters.

Equation (\ref{eq:GFL}) is valid for the interactions considered in the Hamiltonian of Eq.(\ref{eq:hamilt}). Its important to point out that different interactions to the ones considered (e.g. a Hund rule coupling in one of the QD) may change the nature of the ground state and the value of the zero-temperature
conductance.\cite{Roch2008,Roch2009,Logan2009,st,RouraBas2010,Koller2005,Mehta05,Zitko2008,Parks2010,Cornaglia2011,Serge2011,Fabrizio2013_3dots}  

\section{Numerical results}\label{sec:NRG}
We now discuss the temperature dependence of the conductance. We first review the main results for a case with a single relevant level on each QD. This problem has been studied using a variety of techniques including the Numerical Renormalization Group\cite{PhysRevB.71.075305,PhysRevB.81.115316} and Functional Renormalization Group.\cite{Meden2006} 
We first consider a symmetric situation where $\Gamma_L=\Gamma_R=\Gamma/2$, $\tilde{\epsilon}_{D\alpha}=0$, $\mathcal{N}_d=\mathcal{N}_D=1$, and $U_{dD}=0$, so that the average charge on each QD is $1$. 
When both QDs are in the Kondo regime ($U_d,U_D \gg \pi\Gamma, t_{dD}$) charge fluctuations on each QD can be eliminated using a Schrieffer-Wolff transformation\cite{Schrieffer1966} and the low energy properties of the system can be described using a Kondo Hamiltonian:
\begin{equation}
	H_K=J_K \mathbf{S}_d\cdot \mathbf{s}_0+J_{dD}\mathbf{S}_d\cdot \mathbf{S}_D + H_{el}
	\label{eq:KondoHam}
\end{equation}
Here, ${\bf S}_{\ell}$ with $\ell = d,D$ are spin operators
associated to the QDs, and ${\bf s}_{0}=\frac{1}{2}
\sum_{s,s^\prime} c^{\dagger}_{0
s}{\mathbf{\sigma}}_{s,s^\prime}c^{}_{0 s^\prime}$ is the electron spin
density on the orbital coupled to QD $d$. 
The coupling constants, to leading order in $H_t$ and $H_V$, are $J_{dD}=8 t_{dD}^2/(U_d+U_D)$ and 
$J_K = 2 \Gamma/ U_d$.

For a sufficiently weak interdot coupling $J_{dD}$ the system presents a two-stage Kondo effect. 
As the temperature is reduced, the spin $1/2$ of the QD coupled to the leads is Kondo screened at a 
temperature $T_K\sim W e^{-1/\rho J_K}$, where $\rho$ is the lead's local density of states at the Fermi 
level and $W$ is a high energy cutoff. The Kondo effect on QD $d$ generates a peak on its spectral density of width $\sim k_B T_K$, at the Fermi energy. This Kondo peak can be associated to a renormalized Fermi liquid of quasiparticles having a density of states $\sim 1/k_BT_K$.\cite{Nozieres1974} Its emergence results in an increase of the conductance through the device as the temperature is lowered.

For $J_{dD}< T_K$ the spin at QD $D$ is screened at a lower temperature \cite{PhysRevB.71.075305}
\begin{equation}
    T_{K}^\star\sim T_K e^{-\frac{\pi T_K}{J_{dD}}}.
	\label{eq:tstar}
\end{equation}
We can interpret this expression as the Kondo screening of the spin-$1/2$ in the side-coupled QD by the renormalized quasiparticles associated to the Kondo effect in QD $d$. 
While the first stage Kondo effect leads to an increase in the conductance of the device, the second stage Kondo effect suppresses the conductance in agreement with Eq. (\ref{eq:GFL}). This suppression of the conductance can be understood as a Fano antiresonance, or interpreted as a blocking effect in the conductance due to the formation of a strong singlet between the spins on the two QDs at low energies.   
For $J_{dD}> T_K$ the spins in the QDs are also locked in a singlet 
not only at low temperatures, but already
for temperatures $T>T_K$. The conductance is small for $T<J_{dD}$ and the Kondo screening does not takes place. Shifting the gate voltage of QD $d$ to modify its occupation to $\sim 0 $ or $\sim 2$ results in a single stage Kondo effect with an effective Kondo coupling that leads to a high conductance regime at low temperatures.\cite{Tanaka2013} 
When the side-coupled QD in either empty or double occupied and QD $d$ is in the Kondo regime, the transport properties are dominated by the Kondo effect on QD $d$. For sufficiently large tunnel coupling $t_{dD}$ the properties of the system can be better understood considering bonding and antibonding states formed between the two QDs  and including effective Coulomb interactions for the hybridized levels and effective couplings to the leads. \cite{PhysRevB.71.075305}
\begin{figure}[tbp]
\includegraphics[width=8.5cm]{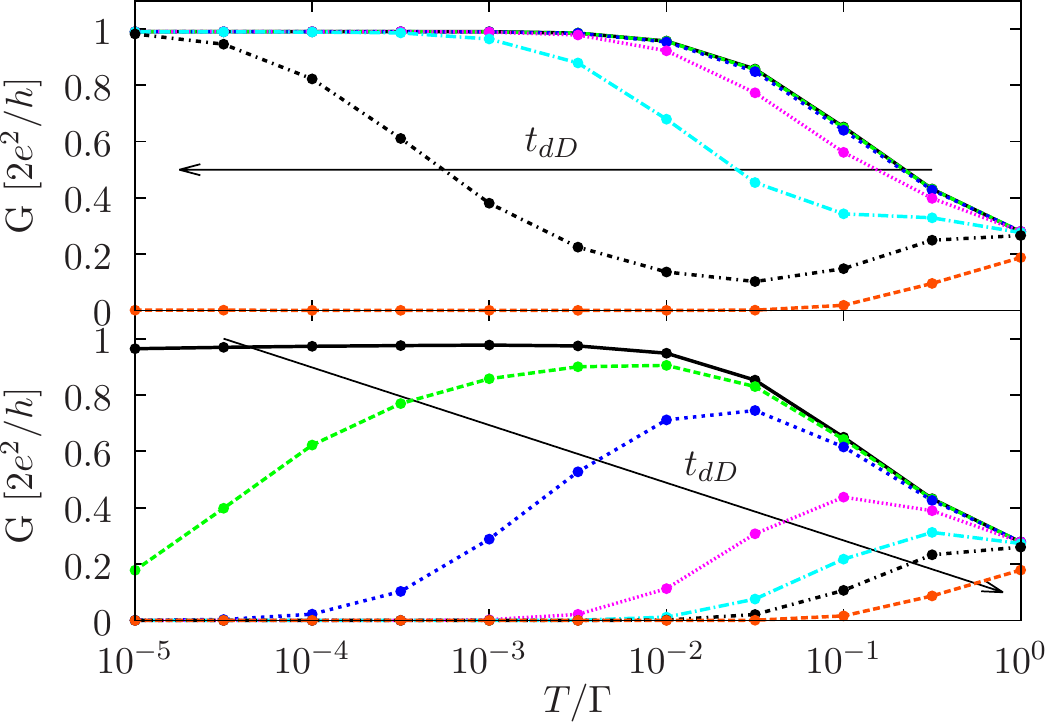}
\caption{ (color online) Conductance for different values of the interdot coupling $t_{dD}=0.01,\, 0.015,\, 0.02,\, 0.03,\, 0.04,\, 0.05$, and $0.09$ meV. The other parameters are $U_{D}=0.25$meV, $U_{d}=0.7$meV, $U_{dD}=0.1$meV, $\delta=0.02$meV, and $\Gamma=0.2$meV. a) Total occupation of tree electrons: $\mathcal{N}_d=1$ and $\mathcal{N}_D=2$. b) Total occupation of four electrons: $\mathcal{N}_d=1$ and $\mathcal{N}_D=3$.  
}
\label{ExpCond}
\end{figure}

As we show below, the main features observed in the single level case for the temperature dependence of the conductance, are also observed when multiple levels are considered in the side-coupled QD.
We calculate the conductance through the system and the magnetic susceptibility using the full density matrix numerical renormalization group (FDM-NRG) \cite{FDM-NRG1, FDM-NRG2}. In all calculations we use a logarithmic discretization parameter $\Lambda=10$ and the $z$-trick \cite{ztrick1, ztrick2} averaging over four values of $z=$ $0.25$, $0.5$, $0.75$, and $1$. We keep up to 5000 states and start the truncation after 4 NRG iterations. 

In Fig. \ref{ExpCond} we present the conductance as a function of the temperature for the multilevel system analyzed in the previous section. On the top panel of Fig. \ref{ExpCond} we consider a situation with a single electron on QD $d$ and $\sim 2$ electrons on QD $D$. For small $t_{dD}\lesssim0.02$meV the results are very similar to what is observed in a decoupled QD situation ($t_{dD}=0$): 
there is an increase 
in the conductance associated to the Kondo screening of the spin-$1/2$ on QD $d$. 
The conductance reaches the quantum of conductance, as expected from the Fermi liquid predictions. 
In this regime, the Kondo temperature is only slightly reduced by the presence of the side-coupled QD.
As $t_{dD}$ is increased there is a decrease of the Kondo temperature which sets the temperature scale at which the Fermi liquid behavior is recovered. This reduction of the Kondo temperature can be understood recalling the analysis of the nature of the electronic wave functions in the different interdot coupling regimes presented in Sec. \ref{sec:wc}. 
For $t_{dD}=0$ two electrons occupy the ground state of QD $D$ forming a singlet and there is a single electron on QD $d$; a situation that remains essentially unaltered for $t_{dD}\ll\delta$. 
For $t_{dD}\gtrsim \delta$, however, it becomes energetically favorable to have a sizable occupation in higher energy levels of QD $D$ in order to increase the hybridization with QD $d$.   
In this regime, the wave-function weight on QD $d$ associated to the spin-$1/2$ is reduced and the Kondo coupling between QD $d$ and the electrodes is reduced accordingly, as it can be readily shown preforming a Schrieffer-Wolff transformation. 
When $t_{dD}$ is increased further, the level structure of the DQD changes (see  also Ref. [\onlinecite{Baines2012}]). Two electrons occupy the lowest lying state of the DQD which is a bonding state between the orbital on QD $d$ and a linear combination of the orbitals on QD $D$. The third electron is mainly localized in QD $D$ which leads to a strongly suppressed Kondo coupling and an exponentially suppressed Kondo temperature. For the highest values of $t_{dD}$ considered, the Kondo temperatures reach values well below the experimentally accessible range (see top panel in Fig. \ref{ExpCond}).

On the bottom panel of Fig. \ref{ExpCond} we consider a situation with a single electron of QD $d$ and $3$ electrons on QD $D$. In this case, the Fermi liquid theory predicts a vanishing conductance at zero temperature. 
The ground state of the isolated DQD has total spin equal to zero. As in the case of a single level in 
QD $D$, if the singlet-triplet gap of the DQD is smaller than the Kondo temperature for QD $d$, 
a two stage Kondo effect occurs as the temperature is decreased. 
First the spin $1/2$ on QD $d$ is Kondo screened by the electrons and holes of the electrodes 
and the low energy excitations of the system can be described by a local Fermi liquid of heavy quasiparticles in QD $d$.
A second stage Kondo effect is observed as the spin $1/2$ on QD $D$ is Kondo screened by the Kondo quasiparticles on QD $d$. The Kondo screening of the side-coupled QD leads to a decrease in the conductance below a characteristic $T_K^\star$ which depends strongly on the antiferromagnetic coupling between the QDs [see Eq. (\ref{eq:tstar})]. 
This behavior can be observed in the bottom panel of Fig. \ref{ExpCond} where for the lowest values of $t_{dD}$ the conductance presents a non-monotonous behavior. As the interdot coupling is increased, the Kondo temperature of the second stage Kondo effect increases exponentially. For large enough values of $t_{dD}$ the antiferromagnetic coupling between the QDs exceeds $T_K$, there is no Kondo effect and the 
conductance decreases monotonously.
While qualitatively the behavior of the conductance is the same as the one observed for a single level on the side-coupled QD, the value of the second stage Kondo temperature can be strongly modified in the multilevel case. As we show in the next section, it can have a strong dependence on the level spacing $\delta$ on QD $D$. 
\begin{figure}[tbp]
\includegraphics[width=8.5cm]{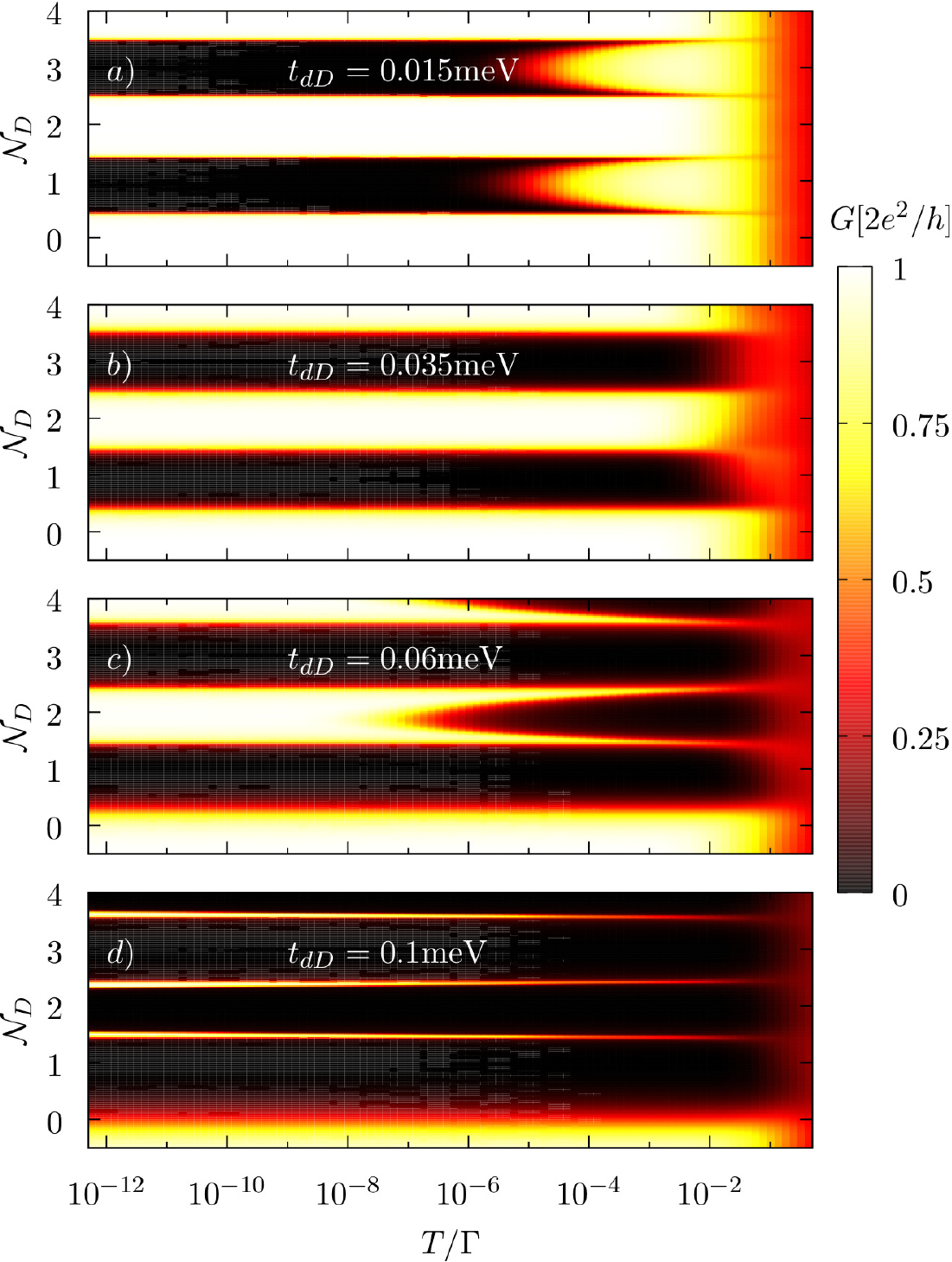}
\caption{(color online) Conductance maps for $\mathcal{N}_d=1$ and different values of $t_{dD}$. Other parameters as in Fig. \ref{ExpCond}. An even-odd asymmetry is observed in the low temperature conductance. }
\label{fig:Cut1}
\end{figure}

\begin{figure}[tbp]
\includegraphics[width=8.5cm]{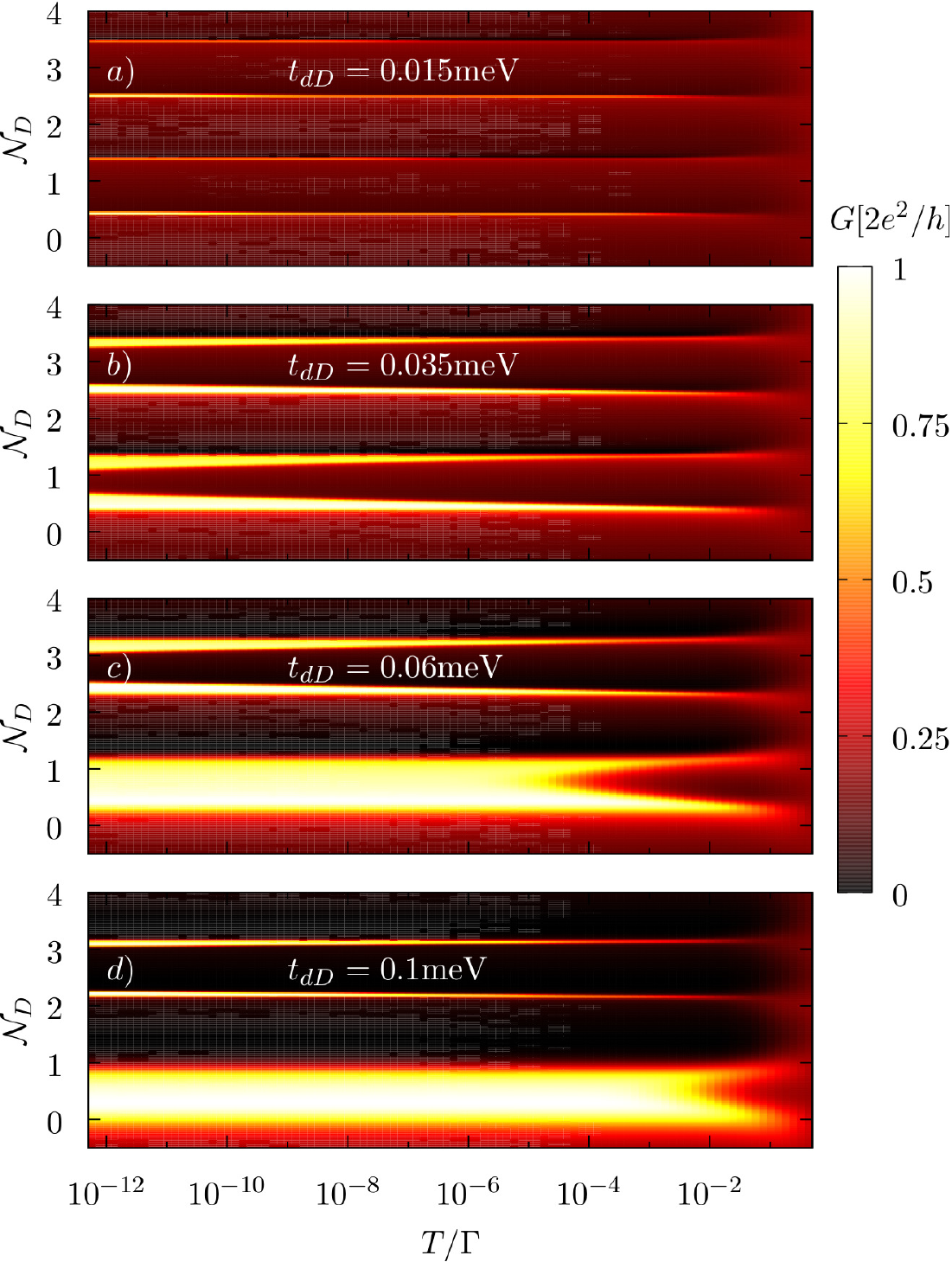}
\caption{ (color online) Same as Fig. \ref{fig:Cut1} for $\mathcal{N}_d=0.2$.}
\label{fig:Cut2}
\end{figure}

The results of Fig. \ref{ExpCond} describe qualitatively the behavior of the conductance away from the charge degeneracy lines in the valleys with an odd or an even number of electrons in the DQD. The temperature scales for the different regimes, however,  can vary strongly from one valley to the other. This is illustrated in Figs. \ref{fig:Cut1} and \ref{fig:Cut2} where the behavior of the conductance as a function of the temperature and $\mathcal{N}_D$ is presented for $\mathcal{N}_d=1$ and $\mathcal{N}_d=0.2$, respectively.
For $\mathcal{N}_d=1$, and a weak interdot coupling $t_{dD}=0.015$meV, a clear even odd asymmetry can be observed in the low temperature conductance as the charge in QD $D$ is modified by the gate voltage. A large (small) conductance is obtained for the valleys with an odd (even) total number of electrons in the DQD. There is, however, an intermediate temperature regime $T_K^\star<T<T_K$ where this even-odd asymmetry is lost. The behavior of the conductance in this parameter regime is qualitatively described by the Hamiltonian of Eq. (\ref{eq:KondoHam}).  

For $t_{dD}=0.035$meV the DQD is in a molecular regime, and the two stage Kondo effect is not observed. 
The system behaves as a single QD that presents the Kondo effect for an odd number of electrons on the DQD, 
and no Kondo physics otherwise. In the Kondo regime a Schrieffer-Wolff transformation can be performed to obtain 
the Kondo coupling that is determined by the effective charging energy of the DQD and the magnetic moment of QD d in the ground-state wave function of the isolated DQD.
For larger values of $t_{dD}$ the Kondo temperature for some of the valleys is strongly suppressed as the orbital on QD $d$ strongly hybridizes with a combination of orbitals on QD $D$ and the remaining wave functions are mainly localized on QD $D$. The odd valley with $\mathcal{N}_D\sim 0$ is associated to the bonding state between the QDs and has therefore a large Kondo temperature. The valleys for $\mathcal{N}_D=2,4$, however, have the magnetic moment mainly localized on QD $D$ leading to a very small Kondo temperature that for $t_{dD}=0.1$meV is well below the temperature range numerically explored (excepting values of $\mathcal{N}_D$ very close to the charge degeneracy points).  

The behavior of the conductance for $\mathcal{N}_d=0.2$ is presented in Fig. \ref{fig:Cut2}. In this case a large conductance is expected at zero temperature for odd values of $\mathcal{N}_D$. For low interdot coupling $t_{dD}=0.015$meV [see Fig. \ref{fig:Cut2}a)]the Kondo temperature is, however, well below the range of temperatures numerically explored. As $t_{dD}$ is increased the Kondo temperature increases and becomes higher than the minimum temperature considered close to the charge degeneracy points [see Fig. \ref{fig:Cut2} b)]. For larger values of $t_{dD}$ [see Figs. \ref{fig:Cut2}c) and d)] the Kondo temperature for $\mathcal{N}_{D} \sim 1$ is strongly enhanced as the interdot bonding state is formed. 
In the valley with $\mathcal{N}_D\simeq 3$, the Kondo temperature reaches a maximum in the molecular regime $t_{dD}\sim\delta$ and becomes strongly suppressed for large $t_{dD}$ as the magnetic moment in the DQD becomes increasingly localized on QD $D$. 

\section{Kondo effect in multilevel Quantum Dots}\label{sec:qdeg}
In this section we analyze the effect of having several electronic levels on QD $D$ on the transport properties of the device and on the Kondo correlations in the two-stage Kondo regime. 
We consider a few levels on QD $D$ and calculate the conductance for the full range of values of the level spacing $\delta$. 
Since we consider a finite number of levels,
the limit of small level spacing $\delta\to 0$ describes finite size QD with a degenerate ground state.
We analyze a DQD system in a parameter regime  which is different to the one analyzed in the previous sections. We set $U_D=U_d=U$, $U_{dD}=0$, and $t_{dD}\ll U_d,U_D$.
Now the Coulomb interactions determine the charge distribution in the DQD, in particular the bonding state obtained in the $t_{dD}>\delta$ regime 
is unfavored by the absence of interdot Coulomb repulsion ($U_{dD}=0$). This can be easily seen calculating the expectation value of the interaction energy term [Eq. \ref{eq:coul}] for a DQD wave-function with a single electron on each QD or the two electrons on the bonding state. 

\begin{figure}[tbp]
\includegraphics[width=8.5 cm]{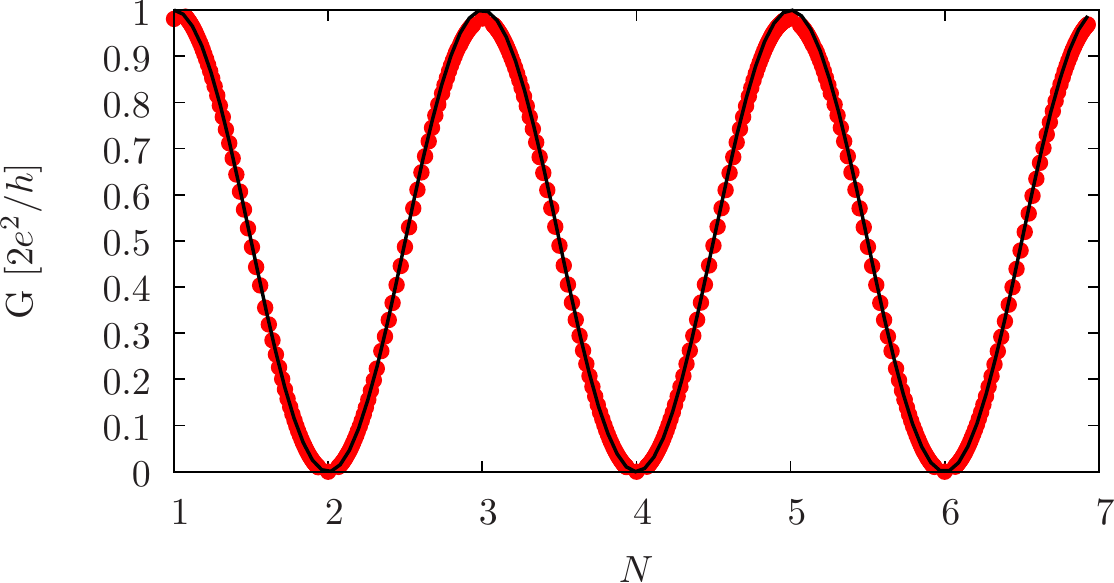}
\caption{(color online) Conductance vs. total DQD occupation for a three level side-coupled QD. NRG (solid dots) and the analytic result (lines) of Eq. (\ref{eq:GFL}) . The gate voltage in the QD $d$ is set such that 
its average occupation is one, while the gate voltage in QD $D$ is varied to change its occupation from zero to six electrons. Parameters 
are $\mathcal{N}_d=1$, $U_{D}/\Gamma=U_{d}/\Gamma=7.143$, $U_{dD}=0$, $t_{dD}/\Gamma=0.0693$, $\delta/\Gamma=0.001428$, and $\Gamma=0.07$ in units of the half-bandwidth of the conduction band of the electrodes.}
\label{Condt485l3Ncx}
\end{figure}

We first analyze the validity of Eq. (\ref{eq:GFL}) for a system having $3$ quasidegenerate levels ($\delta\ll t_{dD}$) on the side-coupled QD calculating the conductance using the Numerical Renormalization Group. We checked that the renormalization procedure had converged to the low energy fixed point to calculate the 
conductance and the occupation. The results are presented in Fig. \ref{Condt485l3Ncx}. The gate voltage of QD $d$ was kept fixed ($\mathcal{N}_d=1$) and the 
occupation on QD $D$ was swept from $0$ to $6$ by changing it's gate voltage. There is an excellent agreement with the Fermi liquid predictions. The small discrepancy between the numerical and analytical results is due to a small loss of spectral weight in the numerical solutions. 
Identical results to those in Fig. \ref{Condt485l3Ncx} were obtained for a wide range of model parameters.  

\begin{figure}[tbp]
\includegraphics[width=8.5cm]{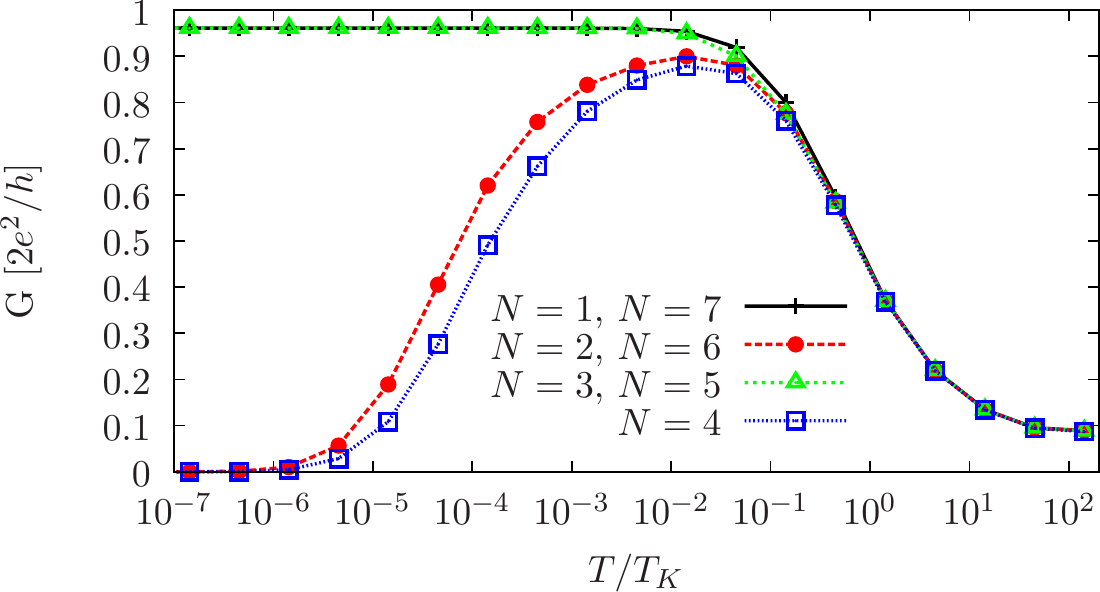}
\caption{(color online) Conductance for different values of the occupation in the DQD. For an even occupation, the conductance is suppressed for  $T\lesssim T_{K}^{\star}$
where the second Kondo stage sets in. Parameters are as in Fig. \ref{Condt485l3Ncx}.}
\label{Cond10t485l3Nx}
\end{figure}

Figure \ref{Cond10t485l3Nx} presents the conductance as a function of the temperature for different values of the total 
DQD occupation and the parameters of Fig. \ref{Condt485l3Ncx}. For an odd number of electrons in the DQD, 
the occupation per spin is $N_\sigma=N/2$ and Eq.(\ref{eq:GFL}) leads to an unitary conductance at zero temperature $G(T=0)=2e^2/h$. The observed increase in the conductance when the temperature is lowered is associated to the buildup of Kondo correlations as the spin in QD $d$ is screened. 
The transport properties are essentially unaltered by the presence of QD $D$ when it is charged by an even number of electrons. 
For an even total number of electrons in the DQD the two stage Kondo effect is obtained, as in the single level case, and the zero-temperature conductance vanishes. The second-stage Kondo temperature generally depends on the number of electrons on QD $D$, 
here is the same for total occupation in the DQD of 
$N$ and $8-N$ electrons due to the electron-hole symmetry for the set of parameters considered.    

We now consider the effect of the level spacing $\delta$ on the second-stage Kondo temperature $T_K^\star$. When there is a single electron on QD $D$, there are two limiting cases that can be readily solved. For $\delta\to\infty$ there is a single relevant level on QD $D$ and the problem reduces to the single level case. For $\delta=0$, we can perform a change of basis on the degenerate ground state of QD $D$, such that $H_t$ couples the level on QD $d$ to only one of the states of the new basis. Namely, the symmetric combination of the original orbitals on QD $D$. 
\begin{equation}
	H_{t}=\sqrt{n_D}t_{dD}\sum_\sigma\left(d^\dagger_{d\sigma} 	\tilde{d}_{D\sigma}+ h.c.\right)
	\label{}
\end{equation}
where $\tilde{d}_{D\sigma}=\sum_{\alpha}	d_{D\alpha\sigma}/\sqrt{n_D}$, and $n_D$ is the degeneracy of the ground state. The problem reduces again to the single level case but with a larger hopping amplitude $\tilde{t}_{dD}=\sqrt{n_D}t_{dD}$. 
Performing Schrieffer-Wolff transformation to the single level problem we obtain $J_{dD}(\delta=0)= n_D J_{dD}(\delta\to\infty)$, leading to a much larger second-stage Kondo temperature in the $\delta=0$ case [$T_K^\star(\delta=0)\gg T_K^\star(\delta\to \infty)$].

We therefore expect to obtain a crossover from a large Kondo temperature to a small Kondo temperature as $\delta$ is increased from zero. Performing a Schrieffer-Wolff transformation in a system with two levels in QD $D$ we have for the Hamiltonian of the isolated DQD to leading order in $H_t$ and $\delta/U$
\begin{eqnarray}
	H_{Dd}&=&J_{1}\mathbf{S}_d\cdot \mathbf{S}_{D1}+J_{2}\mathbf{S}_d\cdot \mathbf{S}_{D2}+\delta\sum_{\sigma}d^\dagger_{D2\sigma}d_{D2\sigma}\nonumber\\
&+&J_{12}\left[\sum_{\sigma}d^\dagger_{D2\sigma}d_{D1\sigma}\mathbf{S}_d\cdot \mathbf{S}_{D1}+h.c.\right]
	\label{eq:KondoHam2}
\end{eqnarray}
with

\begin{eqnarray} \label{Js}
	J_{1}&=&4\frac{t_{dD}^{2}}{U}\nonumber\\
  J_{2}&=&J_{1}\left(1+\frac{\delta^{2}}{U^{2}}\right)\\
  J_{12}&=&\frac{J_{1}+J_{2}}{2}\nonumber
\end{eqnarray}
where, ${\bf S}_{D\ell}$ with $\ell = 1,2$ are spin operators associated to the levels of QD D.

As we show below, the crossover from a large Kondo temperature to a small Kondo temperature does not occur at $\delta\sim J_1$, as it might be naively inferred from an analysis of the isolated DQD, but at a much smaller energy scale $\delta\sim T_K^\star(\delta=0)$.

\begin{figure}[tp]
\includegraphics[width=8.5cm]{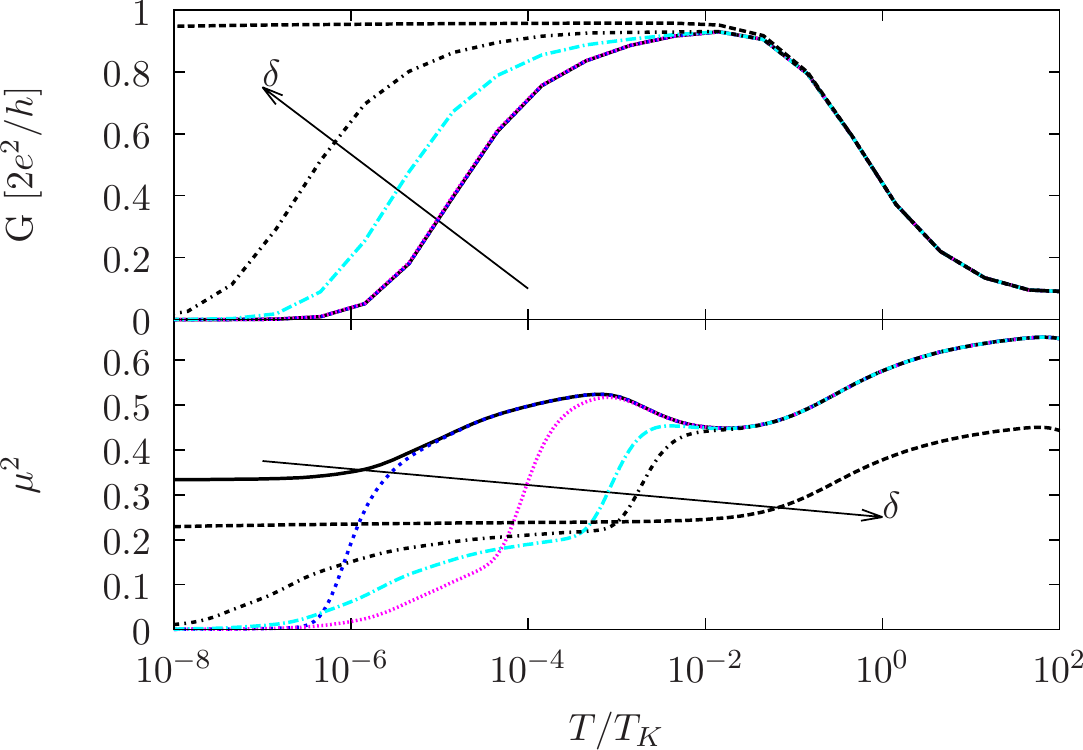}
\caption{(color online) Conductance (top panel) and effective magnetic moment squared (bottom panel) as a function of the temperature for different values of the energy level spacing $\delta/T_K^\star(\delta=0)=0,0.004, 0.4, 3.6, 7.2$, and $\delta\to \infty$ on QD $D$. Other parameters as in Fig \ref{Condt485l3Ncx}. Only for $\delta>T_K^\star(\delta=0)$ there is a sizable change in $T_K^\star$. }
\label{SusCondt3l3Nc3_dx}
\end{figure}
In Fig. \ref{SusCondt3l3Nc3_dx} we present the conductance and the magnetic moment squared $\mu^2$ of the DQD as a function of the temperature for different values of the energy level spacing in QD $D$. The parameters are the same as in Fig. \ref{Cond10t485l3Nx} and the number of electrons in the large dot is odd ($3$). 
The system presents a two stage-Kondo effect as it can be observed in the figure. The conductance increases at temperatures below the first-stage Kondo temperature $T_K$ which is essentially independent of $\delta$.  The second stage-Kondo temperature $T_{K}^\star$ decreases as $\delta$ is increased but only when $\delta$ exceeds $T_{K}^\star(\delta=0)$. Note that for $\delta \to \infty$, the second stage Kondo temperature falls below the minimum temperature numerically explored.

The behavior of $\mu^2$ as a function of the temperature shows a contribution of the three electrons on QD $D$ and suggests that a single electron in an effective orbital couples antiferromagnetically to QD $d$ and is Kondo screened\cite{Boese2002} while the remaining two electrons are doubly occupied and effectively decoupled from the rest of the system. For $T\gg\delta$ these electrons levels contribute with $\mu^2_{D2}=1/3$ to the magnetic moment squared of the DQD and with zero for $T\ll \delta$ presenting a fast crossover between these two regimes as a function of the temperature in contrast to the slow Kondo screening.

As it was mentioned in the previous section, in the second-stage Kondo regime, 
one can consider that
the magnetic moment on QD $D$ couples to a renormalized Fermi liquid of quasiparticles
formed by the Kondo screening of the magnetic moment of QD d. 
In the next section we assume that the first-stage Kondo effect is well developed and explore the crossover of $T_K^\star$ as a function of $\delta$ using a slave-boson theory in the saddle point approximation.

\subsection{Slave-boson mean-field theory}\label{app:SB}

\begin{figure}[tbp]
\includegraphics[width=8.5cm]{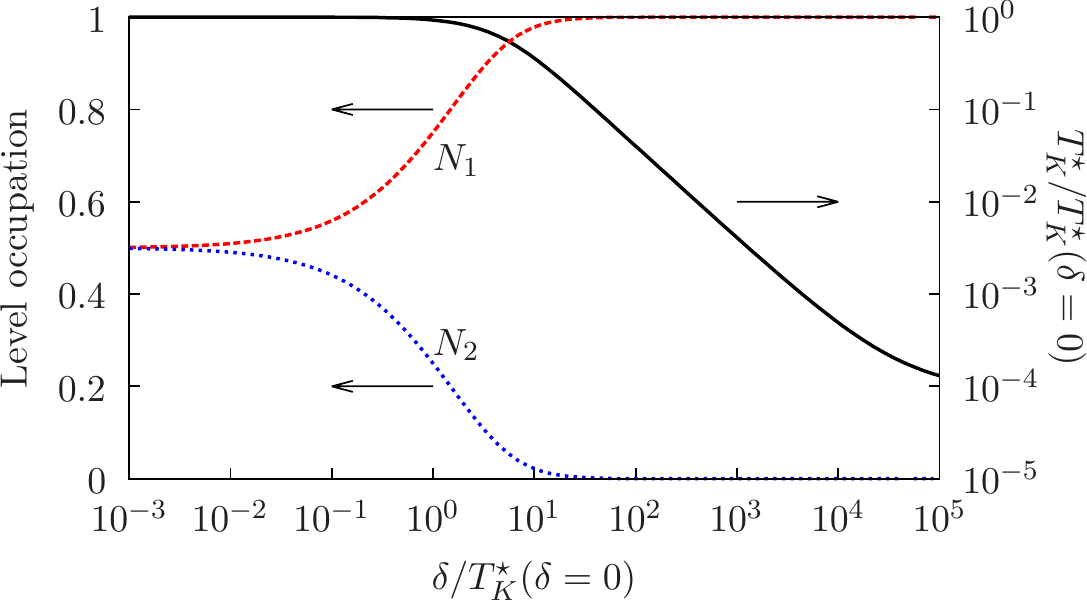}
\caption{(color online) Slave-boson mean-field theory results for the level occupation and the Kondo temperature as a function of the energy level spacing $\delta$ in QD $D$. For $\delta$ larger than $T_K(\delta=0)$ , it becomes energetically favorable to form the Kondo correlations with the lowest lying energy level. This reduces the exchange coupling with the electron bath and the Kondo temperature.}
\label{Bosones}
\end{figure}

Our starting point is the Hamiltonian of Eq. (\ref{eq:KondoHam2}) for a system with two levels on QD $D$ and a 
single electron on each QD. We assume that the first stage Kondo effect is well developed an take the local 
density of states on QD $d$ as that of a non-interacting band with a density of states 
$\rho_\text{eff}\sim1/\pi k_B T_K$ and a bandwidth $\pi k_BT_K$.

For simplicity we consider $\delta \ll U$ and take $J_{1}=J_{2}=J_{12}=J$. The resulting Hamiltonian is:
\begin{eqnarray}
	H_{ml}&=&J \left[\mathbf{S}_D\cdot \mathbf{s}_0^\prime+\sum_{\sigma}d^\dagger_{D2\sigma}d_{D1\sigma}\mathbf{S}_{D1}\cdot \mathbf{s}_0^\prime\right]\\&+&\delta d^\dagger_{D 2\sigma} d_{D 2\sigma}+H_{el}^\prime,
	\label{}
\end{eqnarray}
where $\mathbf{S}_D=\mathbf{S}_{D1}+\mathbf{S}_{D2}$ and ${\bf s}_{0}^\prime=\frac{1}{2}\sum_{s,s^\prime} f^{\dagger}_{0s}{\mathbf{\sigma}}_{s,s^\prime}f^{}_{0 s^\prime}$ is the electron spin
density of the quasi-particles of the Fermi liquid where $f^{\dagger}_{0s}$ $(f_{0s})$ creates (destroys) a quasi-particle on QD $d$, and
\begin{equation}
	H_{\text el}^\prime = \sum_{\nu, k,\sigma} \epsilon_k^\prime f_{\nu k\sigma}^\dagger f_{\nu k\sigma}.
\end{equation}

The biquadratic interactions between the pseudofermions generated by the spin-spin interactions are decoupled, introducing two Bose fields
$B_{i}$ (conjugate to the amplitude $\sum_{\sigma}d^\dagger_{Di\sigma}f_{0\sigma}$ with $i={1,2}$) and the constraint on the occupation of 
QD $D$ is enforced introducing the Lagrange multiplier $\lambda$. The free 
energy expressed in terms of the Bose fields has a saddle point at which the latter condense, 
$\langle B_{i}\rangle=\langle B^\dagger_{i}\rangle=b_{i}=\langle f_{0\sigma}d^\dagger_{Di\sigma}\rangle$. 
At the saddle point the effective Hamiltonian is

\begin{eqnarray}
\widetilde{H}&=&\widetilde{J}\sum_{\sigma,\alpha}
\left(d^\dagger_{D\alpha\sigma} f_{0\sigma} + h.c\right)+ \delta d^\dagger_{D 2\sigma} d_{D 2\sigma} 
\\&+&\sum_{\sigma,\alpha}\lambda (d^\dagger_{D\alpha\sigma} d_{D\alpha\sigma}-1)+ H_{\text el}^\prime,\nonumber
\end{eqnarray}
where $\widetilde{J}=J\left(b_{1}+b_{2}\right)$. Using the equations of motion of the operators $d_{Di\sigma}$ we obtain:

\begin{equation}
    b_{1}=-\frac{1}{\pi}\int_{-T_K}^{T_K} d\omega f(\omega) \text{Im}\left[ \widetilde{\bf G}_{01\sigma}\right],
\label{b1}
\end{equation}

\begin{equation}
    b_{2}=-\frac{1}{\pi}\int_{-T_K}^{T_K} d\omega f(\omega) \text{Im}\left[ \widetilde{\bf G}_{02\sigma}\right],
\label{b2}
\end{equation}

\begin{equation}
    N_{i}=-\frac{2}{\pi}\sum_\sigma\int_{-\infty}^{\infty} d\omega f(\omega) \text{Im}\left[\widetilde{\bf G}_{ii\sigma}\right],
\label{occup}
\end{equation}
where $N_{i}=\sum_{\sigma} \langle d^\dagger_{D\alpha\sigma} d_{D\alpha\sigma} \rangle $ is occupancy of the different levels on QD 
$D$, $\widetilde{\bf G}_{ii\sigma}$ and $\widetilde{\bf G}_{0i\sigma}$ are the Green's function at the $i$-th level of QD $D$ and the 
correlator of the quasiparticles with each level of QD $D$ respectively. These are given by

\begin{equation}
\widetilde{\mathbf{G}}_\sigma=\left(
\begin{array}{ccc}
  \frac{i}{\pi \rho_\text{eff}} & \widetilde{J} & \widetilde{J} \\
    \widetilde{J} & \omega -\lambda  & 0 \\
 \widetilde{J} & 0 & \omega -\delta -\lambda
\end{array}
\right)^{-1}.
    \label{green}
\end{equation}

The integrals in Eq. (\ref{b1}), Eq. (\ref{b2}), and Eq. (\ref{occup}) can be solved analytically at $T=0$. The resulting equations were solved numerically for the level occupations in QD $D$ and the Kondo temperature $T^\star_{k}=\widetilde{\Delta}=\pi\rho_{\text{eff}}\widetilde{J}^2$.
The results are presented in Fig. \ref{Bosones}. We observe a slow crossover from a large Kondo temperature $T^\star_{K}(\delta=0) =\frac{T_K}{2}e^{-\frac{\pi T_K}{2J}}$ to a small Kondo temperature  $T^\star_{K}(\delta\to \infty) =\frac{T_K}{2}e^{-\frac{\pi  T_K}{J}}$. The slow crossover sets in for $\delta \gtrsim T^\star_{K}(\delta=0)$ where the level occupations $N_{1}$ and $N_{2}$ begin to differentiate. For large $\delta\gg T_K^\star(\delta=0)$ only the lowest lying level is occupied. 

If we associate an energy gain $\sim T_K^\star$ to the formation of the Kondo effect, for small $\delta$ it is energetically favorable to 
occupy partially the highest energy level on QD $D$ forming a symmetric combination of the two levels in QD $D$ in order to increase the hybridization with QD $d$ and the Kondo temperature. The energy in this case is $\sim -T_K^\star(\delta=0) +\delta/2$. For $\delta > 2 T_K^\star(\delta=0)$ it is however energetically favorable to empty the highest energy level ($N_2\to 0$) at the expense of having lower energy gain due to the Kondo effect $\sim -T_K^\star(\delta \to\infty)$.
We therefore expect a crossover to occur between the two regimes for $\delta \sim T_K^\star(\delta=0)$.
\section{Conclusions}

We analyzed the electronic transport through a DQD device in the side-coupled configuration. A small quantum dot having a single relevant electronic level is tunnel-coupled to source and drain electrodes, a larger QD is coupled to the small QD but not directly to the electrodes (side-coupled QD). We considered multiple levels on the side-coupled QD and Coulomb interactions between electrons in the DQD. Assuming a Fermi liquid ground state, we obtained an exact relation between the zero-temperature conductance and the electronic occupation of the DQD. Additional interaction terms in the system Hamiltonian, as a Hund's rule 
coupling,\cite{Roch2008,Roch2009,Logan2009,st,RouraBas2010,Koller2005,Mehta05,Zitko2008,Parks2010,Cornaglia2011,Serge2011,Fabrizio2013_3dots} 
may lead to non-Fermi liquid ground states and to a modified zero-temperature conductance formula. \cite{Logan2009}

We explored numerically the conductance through the system for model parameters appropriate to describe two experimentally relevant situations: {\it i)} a system with a large level spacing on the side-coupled QD and {\it ii)} a system with up to three quasidegenerate levels in the side-coupled QD. In the latter case we considered that the quasidegenerate levels were the only relevant for the electronic transport, as it is expected if the remaining levels in the side-coupled QD have a much larger energy. In the former case, which describes the parameter regime of Ref. [\onlinecite{Baines2012}], only a few levels in the side-coupled QD need to be considered to obtain a qualitative description in the weak electrodes-QD coupling regime.

We analyzed the low-temperature conductance and the Kondo correlations for the interdot tunnel-coupling crossover observed in Ref. [\onlinecite{Baines2012}]. Depending on the parity of the number of electrons in the DQD the system may present a two-stage Kondo effect, a single stage Kondo effect or no Kondo effect. We confirmed the predictions of the Fermi liquid theory for $T\to 0$ and found that the temperature at which the Fermi liquid behavior is recovered can be extremely small depending on the model parameters, in particular the tunnel-coupling between the QDs and the gate-voltage on each QD. This leads to conductance maps with an unusual structure at finite temperatures, in the strong interdot coupling regime.

We analyzed the effect of the multilevel nature of the side-coupled QD on the two-stage Kondo effect regime where two Kondo screenings take place in succession as the temperature is lowered. The Kondo temperature of the second stage screening depends strongly on the level spacing of the side coupled QD and the tunnel coupling between the QDs. We considered a system with quasidegenerate levels on the side-coupled QD and constructed an effective model to describe the second stage Kondo effect. We analyzed the resulting model using the slave-boson mean-field approximation. In agreement with the numerical results, we obtained a crossover from a large Kondo temperature for degenerate levels in the side-coupled QD to a small Kondo temperature when the level spacing in the side-coupled QD is of the order of the large Kondo temperature.  In the two stage Kondo regime, the magnetic moment on the side-coupled QD couples to a single electronic channel described by a Fermi liquid of heavy quasiparticles. This multilevel Kondo effect occurs in the presence of a single electronic channel of electrons\cite{Boese2002} instead of multiple channels as would be generally expected for a single multilevel QD coupled to metallic electrodes.

Our results generalize those of Ref. [\onlinecite{PhysRevB.71.075305}] to the multilevel case. The numerical analysis is however restricted to situations where a few electronic levels are relevant in the side-coupled QD. If is of interest to explore the regime where there are many quasidegenerate levels on the side-coupled QD such that Kondo correlations may develop between the two QDs at a temperature larger or of the order of the Kondo temperature of the small QD with the electrodes. In this regime the system may present, at intermediate temperatures, two-channel Kondo physics, \cite{Potok2007} or Kondo box physics if the level spacing on the side-coupled QD is of the order of the energy scale for the Kondo correlations.\cite{PhysRevB.73.205325,PhysRevLett.82.2143,Hu2001,Simon2002,Cornaglia2003b,*Cornaglia2002a,mirage,Kaul2005,Yoo2005,Bomze2010,Kaul2006}
   
\begin{acknowledgments}
We thank Y. Baines, C. Balseiro, D. Feinberg, S. Florens, T. Meunier, and G. Usaj for useful discussions.
We acknowledge financial support from PIP 11220080101821 of CONICET and PICT-Bicentenario 2010-1060 of the ANPCyT.
\end{acknowledgments}

\appendix
\section{Friedel's sum rule}\label{sec:SFR}

In this appendix, we generalize Friedel's sum rule to our system. 
We consider the side-coupled QD with a single level in a small QD and an arbitrary number of levels in the large quantum dot.
For simplicity, we assume that the couplings of the small dot to the leads $\Gamma_L$, $\Gamma_R$ do not depend on the energy.
This usual assumption is justified by the fact that the electronic structure of the leads varies in energy scales much 
larger than both $\Gamma_\nu$.
  
The local Green's function of the DQD is given by:
\be
{\bf G}^{-1}_\sigma(\om)={\bf G}^{-1}_{0\sigma}(\om)-{\bf \Sigma}_\sigma(\om)
\ee
where all the effects of the interactions are included in the self-energy $\Sigma_\sigma(\omega)$
The non-interacting Green's function $G^{-1}_{0\sigma}(\om)$ is given by:
\be
{\bf G}^{-1}_{0\sigma}(\om)= 
\begin{pmatrix} 
\om - \eps_d + i\Gamma/2 & t_1  &t_2  &\cdots & t_N\\ 
t_1 & \om -\eps_1 & 0& \cdots &0\\
t_2 & 0 & \om -\eps_2 &  &\vdots\\
\vdots & \vdots &   &  \ddots &0 \\ 
t_N & 0 & \cdots & 0&  \om-\eps_N\\
\end{pmatrix}
\ee
where $\epsilon_d$ and the $\epsilon_i$ are the single electron energies associated to the levels in QD $d$ and QD $D$, respectively, and include contributions from Eqs. (\ref{eq:coul}) and (\ref{eq:single}). 

The total charge per spin in the DQD is given at $T=0$ by:
\be
N_\sigma=-\frac{1}{\pi} \text{Im} \text{Tr}\int_{-\infty}^0 {\bf G}_\sigma(\om) d\om
\ee
an expression that can be rewritten in the form
\begin{eqnarray}
    N_{\sigma} &= -\frac{1}{\pi}\;\textrm{Im}\;
    \int_{-\infty}^0\;d\omega\; \frac{\partial}{\partial
\omega}\textrm{Tr}\ln {\bf G}^{-1}(\omega) \nonumber\\
&{}- \frac{1}{\pi}\;\text{Im} \int_{-\infty}^0\;d\omega\;
\textrm{Tr} \left[ {\bf G}(\omega) \frac{\partial}{\partial
\omega}{\bf \Sigma}(\omega)\right]\;,
\label{intermediate}
\end{eqnarray}

using the equality
\begin{equation}
    \textrm{Tr} {\bf G}(\omega) = \frac{\partial}{\partial
\omega}\;\textrm{Tr}\ln {\bf G}^{-1}(\omega) +  \textrm{Tr}\left[
{\bf G}(\omega) \frac{\partial}{\partial  \omega}{\bf
\Sigma}(\omega)\right]\;, \label{equivalence}
\end{equation}

The second integral on the right-hand side of
Eq.~(\ref{intermediate}) vanishes order by order in perturbation
theory in the local Coulomb interactions~\cite{Abrikosov} which leads to
\begin{eqnarray}
    N_{\sigma}= \frac{1}{\pi}
\left[\varphi(-\infty)-\varphi(0)\right]\;,
\end{eqnarray}
where
\be
\varphi(\omega) =\text{Tr} \ln \left[ {\bf G}^{-1}(\omega)\right]= \ln \det\left[ {\bf G}^{-1}(\omega)\right].
\label{fase}
\ee
As ${\bf \Sigma}(\om)\lim_{\omega\to 0} \to cte$, the real part of $\det\left[ {\bf G}^{-1}(x)\right]$ diverges as $\om^{N+1}$ while the imaginary part diverges as $\om^N$ resulting in 
\be
\lim_{\omega\to \infty }\text{Im} \ln \det\left[ {\bf G}^{-1}(\omega)\right] = c \times \pi  
\ee
where $c$ is an integer.
Assuming a Fermi liquid ground state, we have $\text{Im}\left[\Sigma(0)\right]=0$ and we obtain
\be \label{eq:Ginvren}
{\bf G}^{-1}_{\sigma}(0)= 
\begin{pmatrix} 
 - \tilde{\eps}_d + i\Gamma/2 & \tilde{t}_{d1}  &\tilde{t}_{d2}  &\cdots & \tilde{t}_{dN}\\ 
\tilde{t}_{1d} & -\tilde{\eps}_1 & \tilde{t}_{12}& \cdots &\tilde{t}_{1N} \\
\tilde{t}_{2d} & \tilde{t}_{12} & -\tilde{\eps}_2 &  &\vdots \\
\vdots & \vdots &   &  \ddots &\tilde{t}_{(N-1)N} \\ 
\tilde{t}_{Nd} & \tilde{t}_{1N} & \cdots & \tilde{t}_{(N-1)N} &  -\tilde{\eps}_N \\
\end{pmatrix}
\ee
\be
N_\sigma=-\frac{1}{\pi}\arctan \left\{\frac{\text{Im}\det\left[{\bf G}^{-1}_\sigma(0)\right] }{\text{Re}\det\left[{\bf G}^{-1}_\sigma(0)\right] }\right\}+c
\ee
Using the Laplacian expansion for the determinant and that the coefficients $\{\tilde{t}_{ij}\}$ and $\{\tilde{\eps}_i\}$ are real we readily find
\be\label{eq:tan2}
\tan^2(\pi N_\sigma)=\left\{\frac{\Gamma D_1/2}{\text{Re}\det\left[{\bf G}^{-1}_\sigma(0)\right] }\right\}^2.
\ee
where $D_1$ is the determinant of the submatrix of ${\bf G}^{-1}_{dd\sigma}(0)$ obtained suppressing the first row and the first column.

The zero temperature conductance per spin through the system is given by:
\be
G_\sigma=\frac{e^2}{\hbar}\frac{\Gamma_R \Gamma_L}{\Gamma_R+\Gamma_L}A_{d\sigma}(0)
\ee
where
\be
A_{d\sigma}(0)= -\frac{1}{\pi}\text{Im}\left[{\bf G}_{dd\sigma}(0)\right].
\ee
Using Cramer's rule for the inverse and Eq. (\ref{eq:Ginvren}) we get
\be
{\bf G}_{dd\sigma}(0)=\frac{D_1}{i \Gamma D_1/2 + \text{Re}\det\left[{\bf G}^{-1}_\sigma(0)\right]}
\ee
and using Eq. (\ref{eq:tan2}) we get
\be
A_{d\sigma}(0)=\frac{2}{\pi \Gamma}\sin^2(\pi N_\sigma)
\ee
and finally 
\be
G_\sigma=\widetilde{\alpha}\frac{e^2}{h}\sin^2(\pi N_\sigma),
\ee
where $\widetilde{\alpha}=\frac{4\Gamma_L \Gamma_R}{(\Gamma_L+\Gamma_R)^2}$.

\bibliographystyle{apsrev4-1}
\bibliography{references}

\end{document}